\documentclass{elsart}
\usepackage{epsfig}
\newcommand{\average}[1]{\left\langle{#1}\right\rangle}
\newcommand{\vs}{\textit{vs.}}
\newcommand{\ie}{\textit{i.e.}}
\newcommand{\eg}{\textit{e.g.}}

\newcommand{\sign}{\mathop\mathrm{sign}}
\newcommand{\pacs}[1]{{\\[0.4cm]\noindent\textit{P.A.C.S.: }{#1}}}
\begin{document}
\begin{frontmatter}
\title{A simple system with two temperatures}
\author[Paris,Naples]{Rapha\"el Exartier\thanksref{Corr}}\ and
\author[Naples,ESPCI]{Luca Peliti\thanksref{INFM}}
\address[Paris]{Laboratoire des Milieux D\'esordonn\'es et 
H\'et\'erog\`enes$^3$\\
Tour 13 - Case 86, 4 place Jussieu, F--75252 Paris Cedex 05 (France)}
\address[Naples]{Dipartimento di Scienze Fisiche,
Universit\`a ``Federico II''\\
Mostra d'Oltremare, Pad.~19, I--80125 Napoli (Italy)}
\address[ESPCI]{Laboratoire de Physico-Chimie Th\'eorique\\
Ecole Sup\'erieure de Physique et Chimie Industrielles\\
10, rue Vauquelin, F--75231 Paris Cedex 05 (France)}
\thanks[Corr]{Author for correspondence. 
E-mail: {\tt exartier@ccr.jussieu.fr}}
\thanks [INFM]{INFM, Unit\`a di Napoli (Italy).}
\thanks[CNRS]{Laboratoire associ\'{e} au CNRS (URA n$^{\circ}$ 800) et 
\`{a} l'Universit\'{e}
Pierre-et-Marie Curie--Paris VI.}

\begin{abstract}
We study the stationary nonequilibrium regime which settles in when two
single-spin paramagnets each in contact with its own thermal bath
are coupled. The response \vs\ correlation plot exhibits some
features of aging systems, in particular the existence, in some
regimes,  of effective temperatures.
\pacs{05.70.Ln Nonequilibrium thermodynamics, irreversible processes}
\end{abstract}
\begin{keyword}
Effective temperatures, fluctuation-dissipation ratio.
\end{keyword}
\end{frontmatter}
A notion of effective temperature for non-equilibrium systems,
first proposed by Hohenberg and Shraiman~\cite{HS} in the
context of turbulent flow, and later developed,
among others, in ref.~\cite{cukupe}, is associated
to the following \textit{Gedankenexperiment}.
A small system (called the \textit{thermometer}) is coupled
both to a heat reservoir at a temperature $\Theta$ and
to the system $\mathcal{S}$ whose effective temperature
$T_{\rm eff}$ must be determined. As a consequence,
heat flows either from or to the reservoir at temperature
$\Theta$. The effective temperature
$T_{\rm eff}$ is defined as the temperature at
which this heat flow vanishes. In aging systems, as well
as in glassy systems kept out of equilibrium by gentle
stirring, it is expected that $T_{\rm eff}$ depends on the
characteristic time scale of the thermomenter, and is related
to the violation of the fluctuation-dissipation theorem.
In order to investigate this phenomenology, one has to
develop a better grasp of the situation in which
a thermodynamical system is in contact
with heat reservoirs at two different temperatures.

Two-temperature systems naturally appear
when considering systems with annealed degrees of freedom, like
magnetic systems with evolving interactions \cite{saakian} 
or diffusing particules on an evolving network \cite{annalisa},
where spins or particules have a temperature different from the one of 
the bonds. We report here the results for what is 
possibly the simplest two-temperature system: two coupled
single-spin paramagnets in contact with different 
thermal baths. When the temperatures of the baths are equal, the 
system reaches a trivial equilibrium state. When the temperatures
are different, the system reaches a
stationary nonequilibrium regime where energy flows through the system
from the high- to the low-temperature thermal bath. 
We analyze this nonequilibrium state by means 
of the susceptibility \vs\ correlation plot~\cite{cuku}. 
At equilibrium, this plot 
is a straight line of slope $-1/T$, as a consequence of the 
fluctuation-dissipation theorem.
In our case, the plot for the high-temperature paramagnet differs 
very slightly from the equilibrium one, but
 its effective temperature~\cite{cukupe} is \textit{smaller} 
than the one of its bath, in contrast with the more common situation.
On the other hand, the low-temperature paramagnet 
can fall deeply out of equilibrium,
and, depending on the parameters of the system, it can violate the 
fluctuation-dissipation theorem in a variety of ways, and
are reminiscent, in some limit, of the corresponding plot
in coarsening systems~\cite{coarsening}.
The ways in which the relation of the cross-response to the cross-correlation
deviate from the equilibrium one are also interesting.

We consider two classical single-spin paramagnets 
coupled via a bilinear spin-spin interaction. 
The Hamiltonian of the system is given by
\begin{equation}
\mathcal{H}=-\half \sum_{i=1,2}r_i S_i^2 +a S_1S_2 ,
\end{equation}
where the ``spin'' variables $S_i$ can take any real value.
The parameters $r_i$ set the response time scale of the paramagnets 
and $a$ represents the strength of the coupling.
Stability requires $a^2< r_1 r_2$. We shall consider $a$ as a small
parameter in the following. 
We recall that the equilibrium correlation $C_i(t)=\average{S_i(t)S_i(0)}$
and response $R_i(t)=\delta\average{S_i(t)}/\delta h_i(0)$ 
of an isolated paramagnet (where $h_i$ is a conjugate
field to $S_i$) are respectively given by
\begin{equation}
C_i(t)=\frac{T}{r_i}\e^{-r_i|t|};\qquad R_i(t)=\theta(t)\e^{-r_i t}.
\end{equation}

The dynamics of the coupled paramagnets is described by  
a set of linear Langevin equations:
\begin{equation}
\partial_t S_i=-\frac{\partial\mathcal{H}}{\partial S_i}+\eta_i(t),
\end{equation}
where $\eta_i$ ($i=1,2$) is a thermal noise, at temperature $T_i$, 
 with zero mean and variance: 
$\langle \eta_i(t) \eta_i(t')\rangle=2 T_i \delta(t-t')$.
Expliciting the previous equation we obtain
\begin{equation}                \label{sys:S,S,para,para}
\left\{
\begin{array}{rcl}
\partial_t S_1&=&-r_1 S_1 +a S_2 + \eta_1(t), \\
\partial_t S_2&=&-r_2 S_2 +a S_1 + \eta_2(t). \\
\end{array}
\right.
\end{equation}

>From this equation we derive a system of eight equations
for the correlation $C_{ij}(t,t')=\average{S_i(t)S_j(t')}$
and response $R_{ij}(t,t')=\delta\average{S_i(t)}/\delta h_j(t')$ 
functions respectively.

For the response functions one has the autonomous equations
\begin{equation}
\left\{
\begin{array}{rcl}
(\partial_t+r_1)\,R_{11}(t,t')&=& a R_{21}(t,t')+\delta(t-t') ,\\
(\partial_t+r_2)\,R_{22}(t,t')&=& a R_{12}(t,t')+\delta(t-t') ,\\
(\partial_t+r_1)\,R_{12}(t,t')&=& a R_{22}(t,t'), \\
(\partial_t+r_2)\,R_{21}(t,t')&=& a R_{11}(t,t') .
\end{array}\right.
\end{equation}
By appying, \eg, $(\partial_t+r_2)$ to the third of these equations,
and substituting the second one, we obtain
\begin{equation}
(\partial_t+r_2)(\partial_t+r_1)\,R_{12}(t,t')=a^2R_{12}(t,t')+a\delta(t-t').
\end{equation}
It is easy to see that $R_{21}$ satisfies the same equation.
Since both $R_{12}$ and $R_{21}$ satisfy the same boundary conditions,
namely they vanish for $t\le t'$, we deduce that 
\begin{equation}
R_{12}(t,t')=R_{21}(t,t'),\qquad \forall\, t,t'.
\end{equation}
For the correlation functions, the equations involve
the response functions:
\begin{equation}          
\left \{
\begin{array}{rcl}
(\partial_t+r_1)\,C_{11}(t,t')&=&aC_{21}(t,t')+2T_1R_{11}(t',t),\\
(\partial_t+r_2)\,C_{22}(t,t')&=&aC_{12}(t,t')+2T_2R_{22}(t',t),\\ 
(\partial_t+r_1)\,C_{12}(t,t')&=&aC_{22}(t,t')+2T_1R_{21}(t',t),\\
(\partial_t+r_2)\,C_{21}(t,t')&=&aC_{11}(t,t')+2T_2R_{12}(t',t) .
\end{array}
\right .
\end{equation} 

After a short transient, the systems enters a stationary regime,
where $C_{ij}(t,t')=\hat C_{ij}(t-t')$ and $R_{ij}(t,t')=\hat R_{ij}(t-t')$.
It is then possible to solve the system.
We define the integrated response $\chi(t)$ by
\begin{equation}
\chi(t)=\int_0^t\d t' \,\hat R(t').
\end{equation}
We then have
\begin{eqnarray}
\label{eq:t,chi_11}
\chi_{11}(t)&=&\theta(t)\left[
-\frac{(r_2-\alpha_{-})(\e^{-\alpha_- t}-1)}{\alpha_{-}(\alpha_{+}-\alpha_{-})}
+\frac{(r_2-\alpha_{+})(\e^{-\alpha_+ t}-1)}{\alpha_{+}(\alpha_{+}-\alpha_{-})}
\right],
 \\
\label{eq:t,C_11}
\hat C_{11}(t)&=& 
\frac{T_1(r_2^2-\alpha_{-}^2)+a^2T_2}{
\alpha_{-}(\alpha_{+}^2-\alpha_{-}^2)} \e^{-\alpha_{-}|t|}
-
\frac{T_1(r_2^2-\alpha_{+}^2)+a^2T_2}{\alpha_{+}(\alpha_{+}^2-\alpha_{-}^2)}
\e^{-\alpha_{+}|t|},
\end{eqnarray}
and the corresponding ones obtained by exchanging the labels 1 and 2. 
We have introduced the following notation for the inverse characteristic
times:
\begin{equation}
\alpha_{\pm}={r_1+r_2 \over 2} \pm {\sqrt {(r_1-r_2)^2+4a^2}\over 2}.
\end{equation}

Since the autocorrelation $C_{11}(t)$ is a monotonically 
decreasing function of $|t|$, we
can invert equation (\ref{eq:t,C_11}) for positive times,
and express the integrated response in terms of the correlation. 
For times longer than $\alpha_+^{-1}$,
when the fastest decreasing exponential in eq.~(\ref{eq:t,C_11}) can be
neglected, we obtain
\begin{equation}
\chi_{11}(t)=\frac{\hat C(0)}{T_1}\left(1-{a^2\over a^2+{r_2(r_1+r_2)
\over T_2/T_1-1}} \right)-\frac{\hat C(t)}{T_1}\left(1-{a^2\over a^2
+{(r_2-\alpha_{-})(r_1+r_2)\over T_2/T_1-1}}\right).\label{eq:CKplot}
\end{equation}
We can directly read off this equation the slope of the response \vs\ 
correlation plot:
\begin{equation}
\frac{X(C)}{T_1}=-\frac{\partial \chi}{\partial C}=
\frac{1}{T_1}\left(1-\frac{a^2}{\displaystyle a^2
+{(r_2-\alpha_{-})(r_1+r_2)\over T_2/T_1-1}}
\right).
\end{equation}
This corresponds to a straight line, \ie, to a well-defined
effective temperature $T_\mathrm{eff}=T_1/X(C)$~\cite{cukupe}.
This temperature goes from $T_1$ to infinity
as $T_2/T_1$ grows from 1 to infinity, showing that the difference from
equilibrium behavior can be very strong in the system coupled
to the colder bath. 

\begin{figure}[htb]
\begin{center}
\epsfig{file=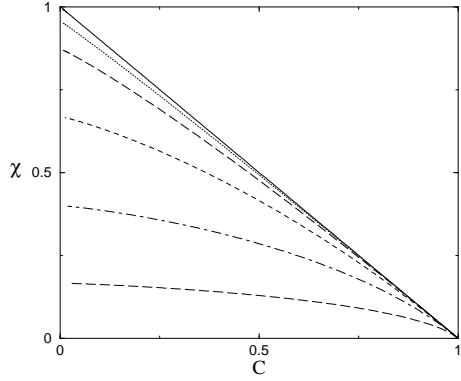,width=6cm}
\end{center}
\caption{Response function $\chi_{11}$ \vs\ correlation function 
$\hat C_{11}$ for $T_1=1$ and
different values of the higher temperature $T_2$.
We have set $r_1=r_2=1$ and $a=0.1$. Both $\chi_{11}$ and
$\hat C_{11}$ are normalized by the instantaneous value $\hat C_{11}(0)$.
The lines correspond (from above to below) to $T_2=1,30,100,300,1000$,
respectively.}\label{ck:fig}
\end{figure}

\begin{figure}[htb]
\begin{center}
\epsfig{file=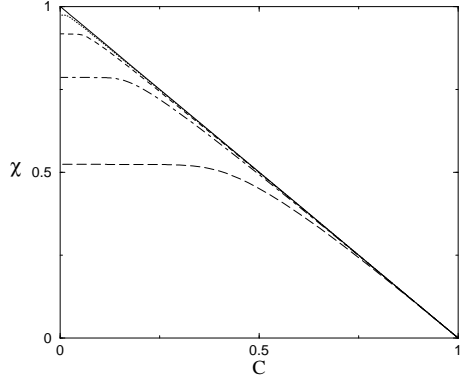,width=6cm}
\end{center}
\caption{Response function $\chi_{11}$ \vs\ correlation function 
$\hat C_{11}$ for $T_1=1$ and
different values of the higher temperature $T_2$.
We have set $r_1=10$, $r_2=1$, and $a=0.1$. Both $\chi_{11}$ and
$\hat C_{11}$ are normalized by the instantaneous value $\hat C_{11}(0)$.
The lines correspond (from above to below) to $T_2=1,100,300,1000,2000$,
respectively. It is interesting to remark that in this
situation the response dies off long before the correlation,
like in coarsening.}\label{ckfast:fig}
\end{figure}

On the other hand,
when $T_2/T_1$ goes from $1$ to $0$, the departure from equilibrium 
is  proportional to $a^2$ and \textit{positive}, \ie, the plot lies
slightly above than the FDT line. 
Thus the high-temperature paramagnet exhibits an
effective temperature slightly \textit{smaller}
than the one of the bath.

The expressions for the cross-correlation and response are:
\begin{eqnarray}
\label{eq:t,chi_12}
\chi_{12}(t)&=&a\,\theta(t)\left[
-{\e^{-\alpha_- t}-1\over\alpha_-(\alpha_+ -\alpha_-)}
+{\e^{-\alpha_+ t}-1\over\alpha_+(\alpha_+ -\alpha_-)}
\right],
 \\
\label{eq:t,C_12}
\hat C_{12}(t)&=&a\left[
{T_1r_2+T_2r_1\over\alpha_+^2-\alpha_-^2}
\left({\e^{-\alpha_-|t|}\over\alpha_-}
-{\e^{-\alpha_+ |t|}\over\alpha_+}\right)\right.\nonumber\\
&&\phantom{a\big[}\qquad\left.{}
-{\sign t\,(T_1-T_2)\over\alpha_+^2-\alpha_-^2}
\left(\e^{-\alpha_-|t|}-\e^{-\alpha_+ |t|}\right)
\right],
\end{eqnarray}
and the corresponding ones obtained by exchanging the labels 1 and 2.
For times longer than $\alpha_+^{-1}$,
when the fastest decreasing exponential in eq.~(\ref{eq:t,C_12}) can be
neglected, the cross-susceptibility has a simple expression in terms of the
cross-correlation:
\begin{eqnarray}
\chi_{12}(t)&=&\frac{r_1+r_2}{r_1T_2+r_2T_1}\hat C_{12}(0)\nonumber\\
&&\phantom{a}-\frac{r_1+r_2}{r_1T_2+r_2T_1}\hat C_{12}(t)
\left(1+{\alpha_-(T_1-T_2)\over r_1T_2+r_2T_1-\alpha_-(T_1-T_2)}\right).
\end{eqnarray}
Notice that the curves of the cross-response \vs\ the cross-correlations
start from the same point (since $\hat C_{12}(0)=\hat C_{21}(0)$, while
$\chi_{12}(0)=\chi_{21}(0)=0$) and end at the same point,
because the symmetry of $\hat R_{ij}(t)$ implies the
equality of $\chi_{12}(t)$ and $\chi_{21}(t)$.
Nevertheless the two \textit{effective} temperatures are different,
one being above and the other below the ``average'' temperature
$\overline T=(r_1T_2+r_2T_1)/(r_1+r_2)$.

\begin{figure}[htb]
\begin{center}
\epsfig{file=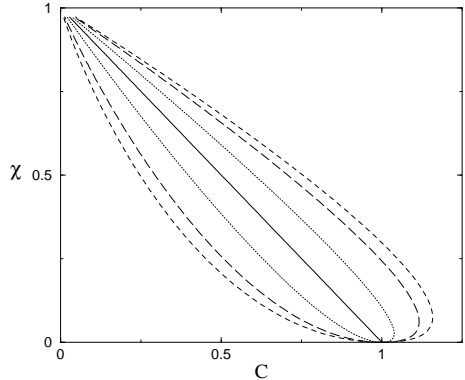,width=6cm}
\end{center}
\caption{Cross-response function $\chi_{12}$ \vs\ correlation functions 
$\hat C_{12}$ and $\hat C_{21}$ for $T_1=1$ and
different values of the higher temperature $T_2$.
We have set $r_1=10$, $r_2=1$, and $a=0.1$.
$\hat C_{12}$, $\hat C_{21}$ are normalized by the instantaneous value 
$\hat C_{12}(0)=\hat C_{21}(0)$, 
and $\chi_{12}$ by its asymptotic value for $t\to\infty$.
The lines correspond (from outside to inside) to $T_2=10,5,2,1$,
respectively. The lines above correspond to $\hat C_{12}$,
those below to $\hat C_{21}$. Notice that the behavior
of $\hat C_{12}$ is not monotonic.}\label{ckcross:fig}
\end{figure}

Summarizing, we have shown that the behavior of a very simple system
with two temperatures exhibits some of the characteristic features of
aging systems, kept out of equilibrium by a stirring force. 
It is possible to extend the approach to study the details of 
the measurement of temperature in an aging system~\cite{prox}.

We thank Prof.~Serge Galam for his interest in this work. L. P. acknowledges
the support of a Chaire Joliot of the E.S.P.C.I.

\end{document}